# SpecGrav – Detection of Gravitational Waves using Deep Learning

Authors: Hrithika Dodia, Himanshu Tandel

Mentor: Lynette D'Mello

Affiliation of all authors and mentor: Dwarkadas J. Sanghvi College of Engineering, Vile Parle, Mumbai, Maharashtra, India

ABSTRACT:

Gravitational waves are ripples in the fabric of space-time that travel at the speed of light. The detection of gravitational waves by LIGO is a major breakthrough in the field of astronomy.

Deep Learning has revolutionized many industries including health care, finance and education. Deep Learning techniques have also been explored for detection of gravitational waves to overcome the drawbacks of traditional matched filtering method. However, in several researches, the training phase of neural network is very time consuming and hardware devices with large memory are required for the task.

In order to reduce the extensive amount of hardware resources and time required in training a neural network for detecting gravitational waves, we made SpecGrav.

We use 2D Convolutional Neural Network and spectrograms of gravitational waves embedded in noise to detect gravitational waves from binary black hole merger and binary neutron star merger. The training phase of our neural network was of about just 19 minutes on a 2GB GPU.

INTRODUCTION:

On 14th September 2015, LIGO detectors at Hanford and Livingston detected gravitational waves for the first time from the merger of two black holes. This event is named GW150914. It is a Nobel Prize winning discovery that substantiates Einstein's general theory of relativity. The sources of gravitational waves detected so far are compact binary coalesces, namely binary neutron star (BNS) merger and binary black hole (BBH) merger. Gravitational waves carry information of their origin with them which includes source masses, spins, distance of the sources from us and much more. They help us understand the objects that are billions of light years far from us, which we could never reach. But when these waves reach us their amplitude can be smaller than the diameter of a proton. Therefore, highly sensitive instruments are required to detect them. To find gravitational waves in detector noise is a meticulous task.

Also, in case of binary neutron star merger, gravitational waves are accompanied by their electromagnetic counterparts. Therefore, rapid detection of gravitational waves in such events is very crucial in order to detect various other remnants of the event like electromagnetic signals and gamma-ray bursts. The method currently used by LIGO for the detection of gravitational waves is Matched Filtering. This method is very time consuming and computationally very expensive. Hence, to overcome the drawbacks of matched filtering many researchers have turned towards deep learning for detection of gravitational waves. Neural networks are trained on a sufficiently large dataset and once trained they can predict output in seconds.

However, one disadvantage of many deep learning techniques used so far for the detection of gravitational waves is that their training phase is very time consuming and requires large memory hardware devices. In this paper, we present a way to reduce the time and resources required for training deep neural network for realtime detection of gravitational waves from binary neutron star (BNS) merger and binary black hole (BBH) merger. We use 2D Convolutional Neutral Network and spectrograms of GW signals embedded in noise for this task. A spectrogram is an image showing the variation of frequency of a signal with time. Using spectrograms instead of time series considerably reduces the size of our dataset.

METHOD:

Our dataset contains spectrograms of signals belonging to three classes:

1. BBH merger
2. BNS merger
3. Noise

To create our dataset, waveforms representing BBH mergers and BNS mergers are generated using LALSUITE by LIGO. For BBH merger signals, masses range from $10M_\odot$ to $36M_\odot$. They are simulated using IMRPhenomD waveform model. We use PhenomDNRT waveform model for simulating BNS merger signals. When dealing with neutron stars, one essential parameter arises – Tidal Deformability. It is the extent till which an object gets distorted by tidal forces. We calculate tidal deformability using the APR equation of state [1]. The source masses for BNS signals lie between $1M_\odot$ and $2M_\odot$.

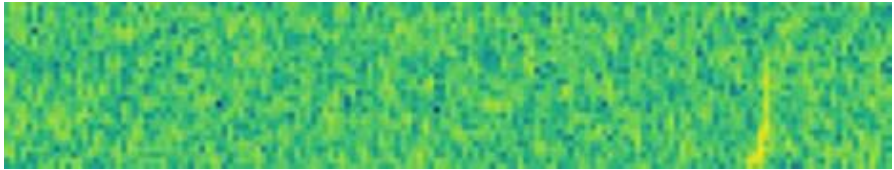

Figure 1: BBH signal injected in real LIGO noise with optimal SNR = 20. The source masses of the merging black holes are $36M_\odot$ and $21M_\odot$.

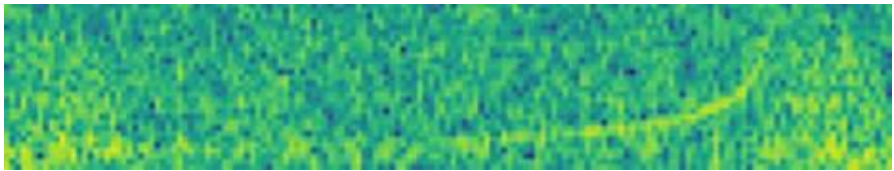

Figure 2: BNS merger signal injected in real LIGO noise with optimal SNR = 32. The source masses of merging neutron stars are $1.8M_\odot$ and $1.1M_\odot$.

We use real LIGO detector noise segments from Hanford detector to create our dataset. The slices of these noise segments are made with an overlap of 4 seconds. Noise and waveforms are whitened separately. The amplitude of waveform is adjusted as per the chosen optimal SNR(Signal to Noise Ratio). This rescaled waveform is injected into noise. Duration of our signal is 5 seconds and sample rate is chosen to be 4096 Hz. Spectrogram is made for each signal containing waveform. For noise class, spectrograms are made for the noise slices from LIGO detector. Our training dataset contains 3500 samples from each class which means total of 10500 samples. Our validation dataset consists of total 1500 samples, containing equal number of samples from each class.

We use 2D Convolutional Neural Network(CNN) for detection of gravitational waves. The input to our CNN is spectrogram of size 256 x 64 x 3. We represent 2D convolution layer as Conv2D(filters, kernel_size), max pooling layer as MaxPooling(pool_size) and fully connected layer as Dense(number_of_units). The architecture of our convolutional neural network can be seen in Figure 3. The activation functions used in our neural network are ReLU(Rectified Linear

Unit) and Softmax. There are 4 convolution layers in our CNN with filters 32, 64, 128 and 128 respectively and kernel size is 3x3 for all of them. For max pooling layers, pool size is 2x2. The strides for the first two max pooling layers is 2x2 and it is 1x1 for rest of them. Whereas, stride is 1x1 for all convolution layers. Dense layers have 128, 64 and 3 units respectively.

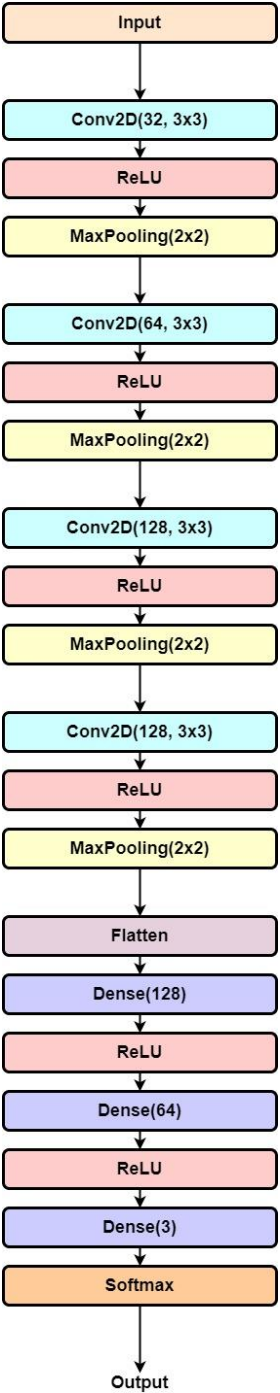

Figure 3: Our convolutional neural network architecture

The optimizer used for training is Adam. Initial learning rate is chosen to be 1e-4 and batch size is 32. We used Tensorflow and Keras to build and train our CNN. We have not used any regularization techniques like BatchNormalization or Dropout in our neural network. Our neural network classifies each input spectrogram as BBH merger signal, BNS merger signal or noise.

RESULT:

We trained our convolutional neural network on a 2GB GPU for just 19 minutes. The requirement of expensive GPUs with high memory to train the neural network is completely eliminated in our method. Also, the time required for completion of the training phase is very less.

For testing the effectiveness of our model in realtime detection of gravitational waves, we evaluate it's performance on real LIGO GW events from GWTC-1. We had first designed a neural network and trained it on dataset containing simulated noise (ZeroDetunedHighPowerNoise). This model had 100% train and validation accuracy but it could not detect most of the real GW events. Our final model (the one presented in this paper) is trained on data containing real LIGO noise and it gives excellent results when tested on real LIGO GW events. We can clearly see the difference between data containing simulated noise and data containing real LIGO noise from the spectrograms below:

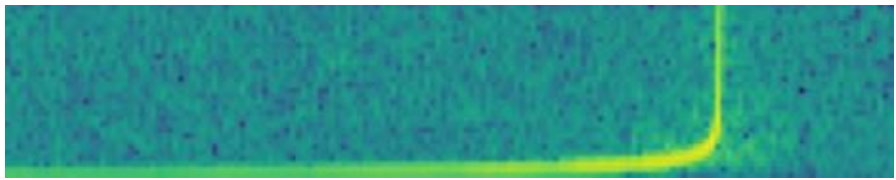

Figure 4: Spectrogram of BBH merger gravitational wave signal injected in simulated noise with optimal SNR = 15

VERSUS

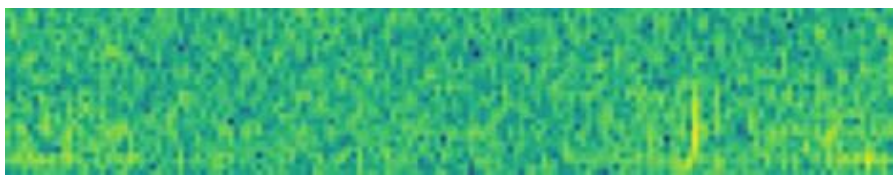

Figure 5: Spectrogram of BBH merger gravitational wave signal injected in real LIGO noise with optimal SNR = 24

Our model correctly identified 9 out of 11 gravitational waves events from GWTC-1. The detailed results can be seen in Table 1.

Table 1: Results of our final model trained with data containing real LIGO detector noise test on real LIGO GWTC-1 events

| NAME OF EVENT | TYPE | RESULT |
|---|---|---|
| GW150914 | BBH MERGER | ✓ |
| GW151012 | BBH MERGER | ✗ |
| GW151226 | BBH MERGER | ✗ |
| GW170104 | BBH MERGER | ✓ |
| GW170608 | BBH MERGER | ✓ |
| GW170729 | BBH MERGER | ✓ |
| GW170809 | BBH MERGER | ✓ |
| GW170814 | BBH MERGER | ✓ |
| GW170817 | BNS MERGER | ✓ |
| GW170818 | BBH MERGER | ✓ |
| GW170823 | BBH MERGER | ✓ |

One of the issues while detection of gravitaional waves are glitches, as they can be easily interpreted as gravitational waves if a detection system is not aware of them. As we use real LIGO noise, many glitches are present in our dataset which made our neural network robust against glitches. For example, the below spectrogram is of noise containing glitch which is correctly classified as noise by our model.

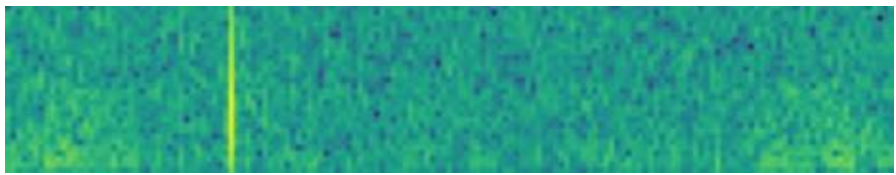

Figure 6: Spectrogram of noise containing glitch

CONCLUSION:

We have successfully demonstrated the use of 2D convolutional neural network for detection of gravitational waves from spectrograms of signals for reducing time required for training the network. Also, minimal hardware was required for training our model. We have trained our model on real LIGO detector noise so that it can be used for realtime detection of gravitational waves.